\documentstyle[12pt,titlepage,epsf]{article}
\def\unlock{\catcode`@=11} 
\def\lock{\catcode`@=12} 
\unlock
  \def\leftrightarrowfill{$\m@th \mathord\leftarrow \mkern-6mu
     \cleaders\hbox{$mkern-2mu \mathord- \mkern-2mu$}\hfill
     \mkern-6mu \mathord\rightarrow$}
  \def\overleftrightarrow#1{\vbox{\ialign{##\crcr
     \leftrightarrowfill\crcr\noalign{\kern-1pt\nointerlineskip}
     $\hfil\displaystyle{#1}\hfil$\crcr}}}
\lock

\newcommand{\bq}{\begin{equation}}
\newcommand{\ba}{\begin{eqnarray}}
\newcommand{\eq}{\end{equation}}
\newcommand{\ea}{\end{eqnarray}}

\def\ddk{[d^3{\rm k}]}
\def\kk{{\rm k}}

\def\b{\beta}

\def\d{\delta}
\def\e{\epsilon}
\def\f{\phi}
\def\vf{\varphi}
\def\g{\raisebox{.4ex}{$\gamma$}}

\def\j{\psi}

\def\m{\mu}
\def\n{\nu}

\def\q{\theta}
\def\r{\rho}

\def\F{\Phi}

\def\J{\Psi}

\def\S{\Sigma}

\def\cc{{\cal C}}
\def\cd{{\cal D}}

\def\ci{{\cal I}}

\def\ct{{\cal T}}

\def\sq{{\lower.2ex\hbox{\large$\Box$}}}

\def\TH{{\raise.2ex\hbox{$\displaystyle \bigodot$}\mskip-4.7mu \llap H \;}}
\def\face{{\raise.2ex\hbox{$\displaystyle \bigodot$}\mskip-2.2mu \llap {$\ddot
        \smile$}}}

\def\Hat#1{\rlap{\kern.10em$\widehat{\phantom G}$}#1}
\def\HAt#1{\rlap{\kern.05em$\widehat{\phantom G}$}#1}

\def\cap#1{\rlap{\kern.1em$\widehat{\phantom{G\vrule height.8em}}$}#1{}}
\def\Cap#1{\rlap{\kern.05em$\widehat{\phantom{G\vrule height.8em}}$}#1{}}

\def\leftrightarrowfill{$\mathsurround=0pt \mathord\leftarrow \mkern-6mu
        \cleaders\hbox{$\mkern-2mu \mathord- \mkern-2mu$}\hfill
        \mkern-6mu \mathord\rightarrow$}
\def\overleftrightarrow#1{\vbox{\ialign{##\crcr
        \leftrightarrowfill\crcr\noalign{\kern-1pt\nointerlineskip}
        $\hfil\displaystyle{#1}\hfil$\crcr}}}

\def\frac#1#2{{\textstyle{#1\over\vphantom2\smash{\raise.20ex
        \hbox{$\scriptstyle{#2}$}}}}}

\catcode`@=11
\def\underline#1{\relax\ifmmode\@@underline#1\else
        $\@@underline{\hbox{#1}}$\relax\fi}
\catcode`@=12

\def\nis{\nointerlineskip}
\def\Abar{\vbox{\nis\moveright.33em\vbox{
        \hrule width.35em height.04em}\nis\kern.05em\hbox{$A$}}{}}
\def\Dbar{\vbox{\nis\moveright.20em\vbox{
        \hrule width.50em height.04em}\nis\kern.05em\hbox{$D$}}{}}
\def\Gbar{\vbox{\nis\moveright.20em\vbox{
        \hrule width.50em height.04em}\nis\kern.05em\hbox{$G$}}{}}
\def\mbar{\vbox{\nis\moveright.15em\vbox{
        \hrule width.60em height.04em}\nis\kern.05em\hbox{$m$}}{}}
\def\Rbar{\vbox{\nis\moveright.20em\vbox{
        \hrule width.50em height.04em}\nis\kern.05em\hbox{$R$}}{}}
\def\Vbar{\vbox{\nis\moveright.05em\vbox{
        \hrule width.60em height.04em}\nis\kern.05em\hbox{$V$}}{}}
\def\Xbar{\vbox{\nis\moveright.20em\vbox{
        \hrule width.60em height.04em}\nis\kern.05em\hbox{$X$}}{}}
\def\thetabar{\vbox{\nis\moveright.15em\vbox{
        \hrule width.30em height.04em}\nis\kern.05em\hbox{$\theta$}}{}}
\def\Lambdabar{\vbox{\nis\moveright.25em\vbox{
        \hrule width.35em height.04em}\nis\kern.05em\hbox{${\mit\Lambda}$}}{}}
\def\Sigmabar{\vbox{\nis\moveright.25em\vbox{
        \hrule width.50em height.04em}\nis\kern.05em\hbox{${\mit\Sigma}$}}{}}
\def\phibar{\vbox{\nis\moveright.18em\vbox{
        \hrule width.40em height.04em}\nis\kern.05em\hbox{$\phi$}}{}}
\def\chibar{\vbox{\nis\moveright.12em\vbox{
        \hrule width.40em height.04em}\nis\kern.05em\hbox{$\chi$}}{}}
\def\psibar{\vbox{\nis\moveright.23em\vbox{
        \hrule width.40em height.04em}\nis\kern.05em\hbox{$\psi$}}{}}
\def\debar{\vbox{\nis\moveright.18em\vbox{
        \hrule width.35em height.04em}\nis\kern.05em\hbox{$\partial$}}{}}
\def\delbar{\vbox{\nis\moveright.10em\vbox{
        \hrule width.63em height.04em}\nis\kern.05em\hbox{$\nabla$}}{}}

\def\rarr{\rightarrow}

\begin{document}
\vspace{1.0in}
{\bf \noindent NONEQUILIBRIUM PROBLEMS IN QUANTUM FIELD THEORY\\[1ex]
 AND SCHWINGER'S CLOSED TIME PATH FORMALISM} \\ \\ \\
\hspace*{1in} Fred Cooper, \\ \\
\hspace*{1in} Theoretical Division, Los Alamos National Laboratory \\
\hspace*{1in} Los Alamos, NM 87545 \\  \\ \\ \\

{\bf  THIS TALK IS DEDICATED TO THE MEMORY OF PROFESSOR JULIAN SCHWINGER
WHO MADE EXTRAORDINARY CONTRIBUTIONS TO THE DEVELOPMENT OF QUANTUM FIELD
THEORY}

\begin{center}
Abstract
\end{center}
 We review the closed time path formalism of Schwinger using a
path integral approach. We apply this formalism to the study of pair
production
from strong external fields as well as the time evolution of a
nonequilibrium
chiral phase transition.

\vspace{3in}
\noindent Talk Presented at the International Conference on Unified
Symmetry in
the Small and in the Large, Miami, Florida.  February 2-5, 1995.

\newpage

{ \bf{INTRODUCTION}} \\ \\

In 1961 in his classic paper ``Brownian Motion of a Quantum Particle"$^1$
, Schwinger solved the formidable technical problem of how to
use the action principle to study initial value problems. Previously, the
action
principle was formulated to study only transition matrix elements from an
earlier time to a later time.  The elegant solution of this problem was the
invention of the closed time path (CTP) formalism. This formalism was
first used to study field theory problems by
Mahanthappa and Bakshi$^1$.

With the advent of supercomputers, it has now become possible to use this
formalism to numerically solve important field theory questions which are
presented as initial value problems.  Two of these problems we shall review
here. They are

\begin{enumerate}
\item {The time evolution of the quark-gluon plasma$^2$.}
\item  {Dynamical evolution of a non-equilibrium chiral phase transition
following a relativistic heavy ion collision$^3$}.
\end{enumerate}

The basic idea of the CTP formalism is to take a diagonal matrix element of the
system at a given time $t=0$ and insert a complete set of states into this
matrix element at a different (later) time $t'$. In this way one can express
the
original fixed time matrix element as a product of transition matrix elements
from $0$ to $t'$ and the time reversed (complex conjugate) matrix element from
$t'$ to $0$. Since each term in this product is a transition matrix element of
the usual or time reversed kind, standard path integral representations for
each
may be introduced. If the same external source operates in the forward
evolution
as the backward one, then the two matrix elements are precisely complex
conjugates of each other, all dependence on the source drops out and nothing
has
been gained. However, if the forward time evolution takes place in the presence
of one source $J_+$ but the reversed time evolution takes place in the presence
of a {\it different} source $J_-$, then the resulting functional is precisely
the generating functional we seek.
\centerline{\epsfysize=3.4in \epsffile{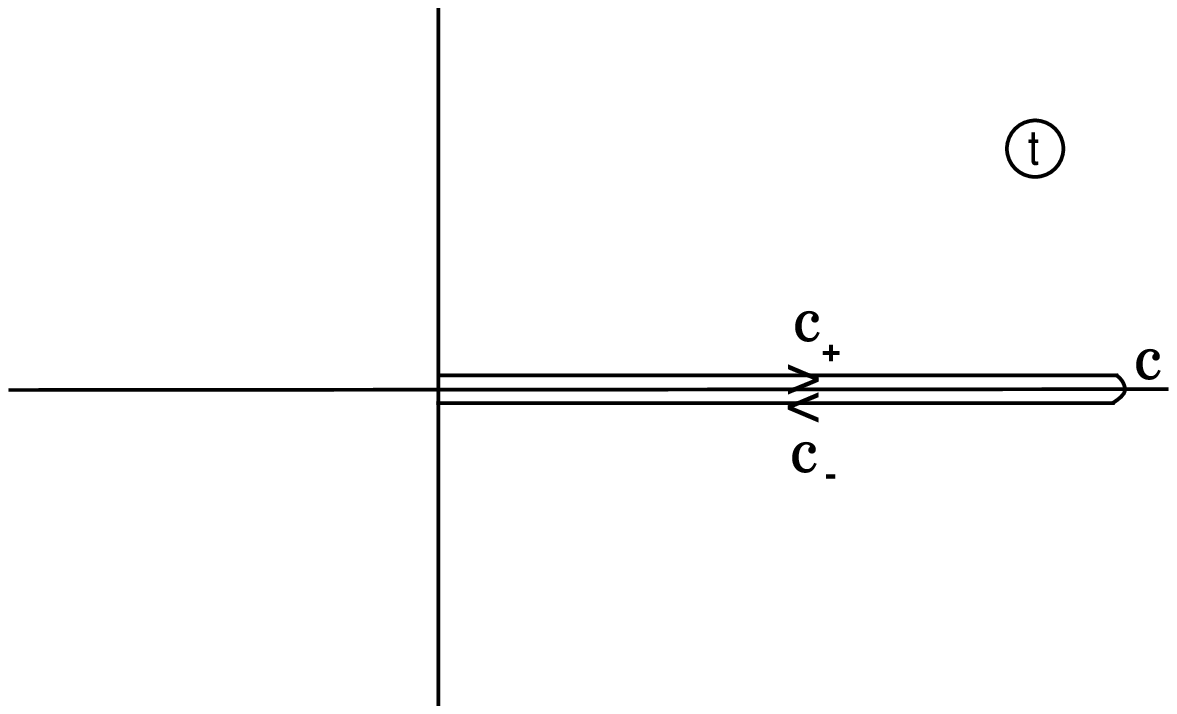}}\vskip4pt\centerline{\bf Fig.
1}
\hangindent\parindent{Complex time contour ${\cal C}$ for the closed time path
propagators.}\\

Indeed
\ba Z_{in}[J_+,J_-] &\equiv& \int [\cd \J]\langle in \vert \j
\rangle_{J_-} \ \langle \j  \vert in \rangle_{J_+} \qquad\qquad\nonumber \\ &=&
\int [\cd \J] \langle in \vert {\ct}^* exp \left[-i \int_{0}^{t^{\prime}} dt
J_{-} (t) \f (t) \right] \vert \J ,t'\rangle \times \nonumber\\ & &
\quad\langle
\J ,t' \vert \ct \exp \left[i \int_{0}^{t^{\prime}} dt J_+ (t) \f (t)\right]
\vert in \rangle \label{eqCTP}  \ea so that, for example, \bq {\d
W_{in}[J_+,J_-]
\over \d J_+(t)}\bigg\vert_{J_+=J_-0} = -{\d W_{in}[J_+,J_-] \over \d
J_-(t)}\bigg\vert_{J_+=J_-=0} =    \langle in \vert \f(t)\vert in \rangle \eq
is
a true  expectation value in the given time-independent Heisenberg state $\vert
in \rangle$. Here $\f(t) = \f(x,t)$ and we are supressing the coordinate
dependence and the integration over the spatial volume in what follows for
notational simplicity.

Since the time ordering in eq. (\ref{eqCTP}) is forward (denoted by $\ct$)
along the time path from $0$ to $t'$ in the second transition matrix element,
but backward
(denoted by $\ct^*$) along the path from $t'$ to $0$ in the first matrix
element, thus the name: {\em{ closed time path generating functional}}.  If we
deform the backward and forward directed segments of the path slightly in
opposite directions in the complex $t$ plane, the symbol $\ct_{\cc}$ may be
introduced for path ordering along the full closed time contour, $\cc$ depicted
in Fig.1.

This deformation of the path corresponds precisely
to opposite $i\e$ prescriptions along the forward and backward directed
segments, which we shall denote by $\cc_{\pm}$ respectively in the following.

If we have an arbitrary initial density matrix $\r$ then we have instead:
 \ba Z \left[ J_{+}, J_{-},\r \right] &\equiv & {\rm Tr} \left\{ \r
\left( \ct^* \exp \left[-i \int_{0}^{t^{\prime}} dt  J_- (t) \f (t)
\right] \right)\left( \ct \exp\left[i \int_{0}^{t^{\prime}} dt J_+ (t) \f(t)
\right] \right) \right\}\qquad\qquad\nonumber \\ &=& \int [\cd \vf]  [\cd
\vf^{\prime}] [\cd \j] \ \langle \vf \vert \r \vert \vf^{\prime}\ \rangle\
\langle \vf^{\prime}\vert \ct^* exp \left[-i \int_{0}^{t^{\prime}} dt
J_- (t) \f (t) \right]\vert \j \rangle   \nonumber \\ & &\quad \times \langle
\j
\vert \ct exp \left[i \int_{0}^{t^{\prime}} dt J_+ (t) \f (t) \right] \vert
\vf\rangle\ . \label{eqZdens} \ea
 Variations of this generating function will
yield Green's functions in the state specified by the initial density matrix,
{\it i.e.} expressions of the form, \bq {\rm Tr}\{ \r  \f(t_{1})  \f(t_{2})
\f(t_{3}) ...\} . \label{green} \eq

Introducing the path integral representation for each transition matrix
element in eq. (\ref{eqZdens}) results in the expression,
\ba
Z \left[ J_{+}, J_{-},\r \right] &=& \int [\cd \vf]  [\cd\vf^{\prime}] \
\langle \vf \vert \r \vert \vf^{\prime} \rangle \int [\cd \j]
\int_{\vf}^{\j} [\cd \f_+]\int_{\vf^{\prime}}^{\j } [\cd \f_-]\
\times \nonumber \\
& & \exp \left[i\int_{0}^{\infty} dt  \left(\ L[\f_{+}] - L[\f_{-}]+J_+\f_+  -
J_-  \f_-  \right) \right]\ , \nonumber \\
\label{eqCTPgen}
\ea
where $L$ is the classical Lagrangian functional, and we have taken the
arbitrary future time at which the time path closes $t' \rarr \infty$.

The double path integral over the fields $\f_+$ and $\f_-$ in (\ref{eqCTPgen})
suggests that we introduce a two component contravariant vector of field
variables by \ba \f^a = \left(\begin{array}{c} \f_+ \\ \f_- \end{array}\right)\
;
\qquad a=1,2
\ea
with a corresponding two component source vector,
\ba
J^a = \left(\begin{array}{c} J_+ \\ J_- \end{array}\right)\ ;
\qquad a=1,2\ .
\ea
Because of the minus signs in the exponent of (\ref{eqCTPgen}), it is necessary
to raise and lower indices in this vector space with a $2 \times 2$ matrix with
indefinite signature, namely \bq c_{ab} = diag \ (+1,-1) = c^{ab} \label{metr}
\eq so that, for example  \bq J^ac_{ab}\F^b = J_+ \f_+ - J_-\f_-\ .  \eq These
definitions imply that the correlation functions of the theory will exhibit a
matrix structure in the $2\times 2$ space. For instance, the matrix of
connected
two point functions in the CTP space is \bq G^{ab}(t,t') = {\d^{2} W \over \d
J_{a}(t) \d J_{b}(t')} \bigg\vert_{J =0}\ . \eq Explicitly, the components of
this $2\times 2$ matrix are   \ba G^{21}(t,t') &\equiv& G_> (t,t') = \ i{\rm
Tr}\{\r\ \F(t) \overline\F(t') \}_{con}\ ,\nonumber \\ G^{12}(t,t') &\equiv&
G_<
(t,t') =  i{\rm Tr}\{\r\ \overline\F(t') \F(t) \}_{con} \\ G^{11}(t,t') &=& \
i{\rm Tr}\left\{\r\ \ct[\F (t) \overline\F(t') ] \right\}_{con} = \q (t,t') G_>
(t,t') + \q (t',t) G_< (t,t')\nonumber \\ G^{22}(t,t') &=& \ i{\rm
Tr}\left\{\r\
\ct^* [\F (t) \overline\F(t')] \right\}_{con}= \q (t',t) G_> (t,t') + \q (t,t')
G_< (t,t') \nonumber  \label{matr} \ea  Notice that  \bq
G^{11}(t,t) = G^{22}(t,t)  \eq with the usual convention that \bq \q (t, t) =
\frac{1}{2}\ . \eq
The $2 \times 2$ matrix notation originated with Schwinger's classic article
in 1960$^1$.

In what follows we will use an alternative generating functional$^4$
 using the Complex Path Ordered Form:
 \ba  \int [\cd \j] \langle \vf^{\prime} \vert \ct^* \exp \left[-i
\int_{0}^{\infty} dt \, J_{-} (t) \f (t) \right] \vert \j \rangle \langle \j
\vert \ct \exp \left[i \int_{0}^{\infty} dt \, J_+ (t) \f(t)\right] \vert \vf
\rangle \qquad\nonumber \\ =\langle \vf^{\prime}  \vert \ct_{\cc} \exp \left[i
\int_{\cc} dt \, J (t) \f(t)\right] \vert \vf \rangle\qquad\qquad\qquad \ea so
that (\ref{eqZdens}) may be rewritten more concisely in the CTP complex path
ordered form, \ba Z_{\cc} \left[ J, \r \right] &= & {\rm Tr} \left\{ \r  \left(
\ct_{\cc} \exp\left[i \int_{\cc} dt J (t) \f (t) \right] \right)
\right\}\qquad\qquad\qquad\nonumber \\ &=& \int [\cd \vf^1] \int [\cd \vf^2]\
\langle \vf^1 \vert \r \vert \vf^2\ \rangle \int_{\vf^1}^{\vf^2} [\cd \f] \exp
\left[i\int_{\cc} dt  \,\left( L[\f] + J\f\right) \right]\ . \label{Zfin} \ea

This is identical in structure to the usual
expression for the generating functional in the more familiar in-out formalism,
Only difference -- path ordering according to the complex time contour
$\cc$ replacing the ordinary time ordering prescription along only $\cc_+$.
 For example, the propagator function becomes \ba G(t,t')
&=&\q_{\cc} (t,t') G_>(t,t') + \q_{\cc} (t',t) G_<(t,t')\nonumber\\
 &\equiv & \q_{\cc} (t,t') G^{21}(t,t') + \q_{\cc} (t',t) G^{12}(t,t')
\label{CTPg} \ea where $\q_{\cc}$ is the CTP complex contour ordered theta
function defined by
\ba \q_{\cc} (t,t') \equiv \left\{ \begin{array}{ll} \q
(t,t') &\mbox{for $t,t'$ both on $\cc_+$}\\ \q (t',t)&\mbox{for $t,t'$ both on
$\cc_-$}\\ $1$ &\mbox{for $t$ on $\cc_-$ , $t'$ on $\cc_+$}\\ $0$ &\mbox{for
$t$
on $\cc_+$ , $t'$ on $\cc_-$}\end{array} \right. \label{eq:theta} \ea  With
this
definition of $G(t,t')$ on the closed time contour, the Feynman rules are the
ordinary ones, and matrix indices are not required. In integrating over the
second half of the contour $\cc_-$ we have only to remember to multiply by an
overall negative sign to take account of the opposite direction of integration,
according to the rule, \bq \int_{\cc}dt = \int_{0\ \cc_+}^{\infty}dt - \int_{0\
\cc_-}^{\infty} dt\ . \label{neg} \eq

A second simplification is possible in the form of the generating functional of
(\ref{Zfin}), if we recognize that it is always possible to express the matrix
elements of the density matrix as an exponential of a polynomial in the
fields$^5$
 \bq \langle \vf^1 \vert \r \vert \vf^2\ \rangle = \exp
\left[R +  R_a(t_0)\vf^a(t_0) +  R_{ab}(t_0)\vf^a(t_0)\vf^b(t_0) +
\ldots\right]\ . \eq Since any density matrix can be expressed in this form,
there is no loss of generality involved in expressing $\r$ as an exponential.
If
we add this exponent to that of the action in (\ref{Zfin}), and integrate over
the two endpoints of the closed time path $\vf^1$ and $\vf^2$, then the only
effect of the non-trivial density matrix $\r$ is to introduce source terms into
the path integral for $Z_{\cc}[J,\r]$ with support {\it only} at the endpoints.
This means that the density matrix can only influence the boundary conditions
on
the path integral at $t=0$, where the various coefficient functions $R_a$,
$R_{ab}$, {\it etc.} have the simple interpretations of initial conditions on
the
one-point (mean field), two-point (propagator), functions {\it etc.} It is
clear
that the equations of motion for $t \ne 0$ are not influenced by the presence
of
these terms at $t_0=0$. In the special case that the initial density matrix
describes a thermal state, $\r_{\b} = \exp\{-\b H\}$ then the trace over
$\r_{\b}$ may be represented as an additional functional integration over
fields
along the purely imaginary contour from $t=-i\b$ to $t=0$ traversed before
$\cc_-$ in Fig. 1. In this way the Feynman rules for real time thermal Green's
functions are obtained$^6$.
 Since we consider general nonequilibrium
initial conditions here we have only the general expression for the initial
$\r$
above and no contour along the negative imaginary axis in Fig. 1.

To summarize, we may take over all the results of the usual scattering theory
generating functionals, effective actions, and equations of motion  provided
only that we
\begin{enumerate} \item substitute the CTP path ordered Green's
function(s) (\ref{CTPg}) for the ordinary Feynman propagators in internal
lines;
\item integrate over the full closed time contour, $\cc$, according to
(\ref{neg}); and  \item satisfy the conditions at $t=0$ corresponding to the
initial density matrix $\r$. \\ \\
\end{enumerate}
{\bf Closed time path contour and causality } \\

 Rules for evaluating the time integrals using the closed time path [CTP]
contour shown in Fig.~(1).

The integration path is given by
\begin{equation}
   \int_{c} {\rm d}t =
     \int_{0: c_+}^{\infty} {\rm d}t -
     \int_{0: c_-}^{\infty} {\rm d}t  \>.
\end{equation}
The {\em causal} Green's functions are given by functions of the form,
\begin{equation}
   A(t,t') = \Theta_{c}(t,t') A_{>}(t,t') +
             \Theta_{c}(t',t) A_{<}(t,t')  \>,
\end{equation}
where $\Theta_{c}(t,t')$ is defined in  eq.(\ref{eq:theta})
 These causal Green's functions are
symmetric.
\[ A_{>}(t,t') = A_{<}(t',t) \]
To prove causality of any graph we need  two lemmas.
\begin{enumerate}
\item{Lemma 1- a loop of two causal functions,such as
self energy graph,
is another causal function}.
\begin{eqnarray}
   B(t,t') & = & \Theta_{c}(t,t') B_{>}(t,t') +
                 \Theta_{c}(t',t) B_{<}(t,t')
   \nonumber \\
   C(t,t') & = & \Theta_{c}(t,t') C_{>}(t,t') +
                 \Theta_{c}(t',t) C_{<}(t,t') \>,
\end{eqnarray}
the self energy graph
$$A(t,t') = i B(t,t') C(t,t'),$$
then
\begin{equation}
A(t,t') = \Theta_{c}(t,t') A_{>}(t,t') +
                 \Theta_{c}(t',t) A_{<}(t,t')  \label{eq:lemma}
\end{equation}
where
\begin{equation}
A_{>,<}(t,t')= i B_{>,<}(t,t')C_{>,<}(t,t')
\end{equation}
\item{Lemma 2 -Matrix product of two causal functions is causal}
\begin{equation}
   A(t_1,t_3) =  \int_{c} {\rm d}t_2 \, B(t_1,t_2) C(t_2,t_3)
   \>,
\end{equation}
we find then
\begin{equation}
   A(t,t') = \Theta_{c}(t,t') A_{>}(t,t') +
             \Theta_{c}(t',t) A_{<}(t,t') \label{eq:lemma2} \>,
\end{equation}
where
\begin{eqnarray}
   A_{ \stackrel{>}{<} }(t_1,t_3) & = &
     - \int_{0}^{t_3} {\rm d}t_2 \, B_{ \stackrel{>}{<} }(t_1,t_2)
     \left[ C_{>}(t_2,t_3) - C_{<}(t_2,t_3) \right]
   \nonumber \\ &&
     + \int_{0}^{t_1} {\rm d}t_2 \,
     \left[ B_{>}(t_1,t_2) - B_{<}(t_1,t_2) \right]
      C_{ \stackrel{>}{<} }(t_2,t_3)  \>.
\label{eq:matrixmult}
\end{eqnarray}

\end{enumerate}
Now consider the product of three causal functions:
\begin{equation}
   A(t_1,t_4) =  \int_{c} {\rm d}t_2  \int_{c} {\rm d}t_3 \,
     B(t_1,t_2) C(t_2,t_3) D(t_3,t_4)  \>.
\end{equation}
We can work this case out by applying the second lemma from left to right.
That is, we can let
\begin{equation}
   E(t_1,t_3) = \int_{c} {\rm d}t_2 \, B(t_1,t_2) C(t_2,t_3)  \>.
\end{equation}
Then $E(t_1,t_3)$ is causal.  We are then left with:
\begin{equation}
   A(t_1,t_4) =  \int_{c} {\rm d}t_3 \, E(t_1,t_3) D(t_3,t_4) \>,
\end{equation}
and so $A$ is also causal.

 After doing the integrals sequentially
one is left with
\begin{equation}
  f(t) = \int_c dt_1 F(t,t_1) = \int_0^{t} [F_{>}(t,t_1) -F_{<}(t,t_1)],
\end{equation}
which explicitly displays the causality (dependence only on earlier times).

To see how these rules work in practice, consider the terms contributing
to order $1/N$ to the induced current determining the backreaction on
an initially strong electric field$^4$. The current is just
$ {\rm tr} \{ \gamma^{\mu} \tilde{G} \} $. Where $\tilde{G}$ is the full
fermion
propagator to order $1/N$.

Using the above lemmas we  obtain
 that the  Maxwell eqs. of motion take the form,
\ba \partial_{\n} F^{\m\n}(x) &=& \langle j^{\m}(x)\rangle = -{ie^2\over 2}{\rm
tr} \left\{\g^{\m}[G_>(x,x) + G_<(x,x)]\right\}\nonumber\\ &+&{2e^2\over N}
\int_0^t dt_1 d^3 \vec x_1 \int_0^{t_1} dt_2 d^3 \vec x_2 \ci m {\rm
tr}\biggl\{\g^{\m}\left[G_>(x,x_1) - G_<(x, x_1)\right] \times\nonumber \\  &&
\quad \left[\S_<(x_1,x_2) G_>(x_2,x) - \S_>(x_1,x_2) G_<(x_2,x) \right]\
\biggr\}\ . \label{max1}  \ea

Here:
\begin{equation}
\S_<(x_1,x_2)=  i \gamma^{\mu} G_<(x_1,x_2) \gamma^{\nu} D_{\nu \mu <
} (x_1,x_2). \end{equation}

This immediately displays the causality of the result: that is only previous
times contribute to the space-time integrations.\\ \\
{\bf TIME EVOLUTION OF THE QUARK GLUON PLASMA} \\\\
Our model for the production of the quark-gluon plasma begins with the creation
of a flux tube containing a strong color electric field. We assume the
kinematics
of ultrarelativistic high energy collisions results in a boost invariant
dynamics
in the longitudinal  ($z$) direction
where all expectation values are functions of the proper time $\tau =
\sqrt{t^2-z^2}$.We
therefore introduce the light cone variables
$\tau$ and $\eta$, which will be identified later with
fluid proper time and rapidity . These coordinates are defined in terms of the
ordinary lab-frame Minkowski time $t$ and coordinate along the beam direction
$z$ by
\begin{equation}
       z= \tau \sinh \eta  \quad ,\quad t= \tau \cosh \eta \,.
\label{boost_tz.tau.eta}
\end{equation}
 The Minkowski line element in these coordinates has the form
\begin{equation}
{ds^2} = {- d \tau^2 + dx^2   +dy^2 +{\tau}^2 {d \eta}^2 }\,.
\label{boost_line_element}
\end{equation}
Hence the metric tensor is given by
\begin{equation}
 g_{\mu \nu} = {\rm diag} (-1, 1, 1, \tau^2).
\end{equation}

For simplicity, here we discuss pair production from an abelian Electric Field
and the subsequent quantum back-reaction on the Electric Field. The physics
of the problem can be understood for constant electric fields as a simple
tunneling process. If the electric field can produce work of at least
twice the rest mass of the pair in one compton wavelength, then the vacuum
is unstable to tunnelling. This condition is:
\bq { eE \hbar \over m c} \ge 2 m c^2.  \eq

The problem of pair production from a constant Electric field (ignoring the
back reaction) was studied by J. Schwinger in
1951 $^7$.
 The WKB argument is as follows:
    One imagines an electron bound by a potential well of order
 $|V_{0} |\approx 2m$ and submitted to an additional electric potential
$eEx$ .  The ionization probability  is proportional to the WKB barrier
penetration factor:

\bq \exp [-2 \int_{o}^{V_{o}/e} dx \lbrace 2m (V_{o} -|eE| x)\rbrace^{1/2}]
= \exp (-{4 \over 3} m^{2}/|eE|)
\eq
In his classic paper Schwinger was able to analytically solve for the effective
Action
in a constant background electric field and determine an exact pair production
rate:
\bq
w = [\alpha E^{2}/(2\pi^{2})] {\sum_{n=1}^{\infty}} {(-1)^{n+1} \over n^{2}}
\exp (-n\pi m^{2}/|eE| ). \eq
By assuming this rate could be used when the Electric field was slowly varying
in time,
the first back reaction calculations were attempted using semi classical
transport methods.
Here we use the $CTP$ formalism and perform the field theory calculation.
The lagrangian density for QED in curvilinear
coordinates
gives rise to the action
\begin{equation}
S= \int d^{d + 1}x \, ({\rm{det}}\, V) \left[ {\frac{-i}{2}}
\bar {\Psi} \tilde{\gamma}^{\mu}
\nabla_{\mu} \Psi+ {\frac{i}{2}} (\nabla^{\dag}_{\mu}\bar {\Psi} )
\tilde{\gamma}^{\mu} \Psi  -i m \bar {\Psi}\Psi-
 {1 \over 4}F_{\mu \nu}F^{\mu \nu} \right],
\label{boost_Sf}
\end{equation}
where$^8$
\begin {equation}
\nabla_{\mu} \Psi \equiv (\partial_{\mu} + \Gamma_{\mu} -ieA_{\mu}) \Psi
\end{equation}

{}From the action (\ref{boost_Sf}) we obtain the Heisenberg field
equation for the fermions,
\begin{equation}
\left( \tilde{\gamma}^{\mu}\nabla_{\mu} + m \right) \Psi=0\,,
\end{equation}
which takes the form

\begin{equation}
\left[ \gamma^0 \left(\partial_\tau+{1 \over 2\tau}\right)
+{\bf \gamma}_\perp\cdot \partial_\perp
+ {\gamma^3 \over \tau}(\partial_\eta -ieA_\eta)+ m \right] \Psi =0\,,
\label{boost_Dirac}
\end{equation}
Variation of the  action with respect to $A_{\nu}$ yields the Maxwell
equations:
If the electric field is in the $z$~direction and a function of $\tau$ only,
we find that the only nontrivial Maxwell equation is
\begin{equation}
{1 \over \tau}  {dE(\tau) \over d\tau} =  {e \over 2} \left \langle \left[
\bar{\Psi}, \tilde {\gamma}^{\eta} \Psi \right] \right \rangle
={e \over 2 \tau} \left \langle \left[ \Psi^{\dagger}, \gamma^0\gamma^3 \Psi
\right] \right \rangle .
\label{boost_MaxD2}
\end{equation}

We expand the fermion field in terms of Fourier modes at fixed proper time
$\tau$,
\begin{equation}
\Psi (x) = \int [d{\bf k}] \sum_{s}[b_{s}({\bf k})
\psi^{+}_{{\bf k}s}(\tau)
 e^{i k \eta} e^{ i {\bf{p}} \cdot {\bf x}}
+d_{s}^{\dagger}({\bf{-k}}) \psi^{-}_{{\bf{-k}}s}(\tau)
e^{-i k \eta} e^{ - i {\bf{p}} \cdot {\bf x}}  ].
\label{boost_fieldD}
\end{equation}
The $\psi^{\pm}_{{\bf k}s}$ then obey
\begin{eqnarray}
 \left[\gamma^{0} \left({d\over d \tau}+{1 \over 2\tau}\right)
+ i \gamma_{\bf{{\perp}}}\cdot{\bf{k_{\perp}}}
+i {\gamma^{3}} \pi_{\eta}
 + m\right]\psi^{\pm}_{{\bf k}s}(\tau) = 0,
\label{boost_mode_eq_D}\end{eqnarray}
 We square the Dirac equation by introducing
\begin{equation}
\psi^{\pm}_{{\bf k}s} = \left[-\gamma^{0}\left( {d \over d\tau}
+{1 \over 2\tau}\right)
- i \gamma_{\bf{{\perp}}}\cdot {\bf{k_{\perp}}}
-i \gamma^{3} \pi_{\eta}+ m\right] \chi_{s} {f^{\pm}_{{\bf k}s} \over {\sqrt
\tau}}
\, . \label{boost_psi_g}
\end{equation}
The spinors $\chi_s$ are chosen to be eigenspinors of $\gamma^0\gamma^3$,
\begin{equation}
\gamma^{0}\gamma^{3}\chi_{s} = \lambda_{s} \chi_{s}\,
\end{equation}
with $\lambda_{s}=1$ for $s=1,2$ and $\lambda_{s}=-1$ for $s=3,4$.
They are normalized,
\begin{equation}
\chi^{\dag}_{r} \chi_{s} = 2\delta_{rs} \,.
\label{boost_e.v}
\end{equation}
The sets $s={1,2}$ and $s={3,4}$ are two different complete sets of
linearly independent solutions of the Dirac equation
Inserting (\ref{boost_psi_g}) into the Dirac equation
(\ref{boost_mode_eq_D})  we obtain the quadratic mode equation
\begin{equation}
\left( {d^2 \over d \tau^2}+
\omega_{\bf k}^2 -i \lambda_{s} \dot{\pi}_{\eta} \right )
f^{\pm}_{{\bf k}s}(\tau) = 0,
\label{boost_modef_D}
\end{equation}
where now
\begin{equation}
 \omega_{\bf k}^2= \pi_{\eta}^2 +{\bf k}_{\perp} ^2 +m^2 .
\label{boost_omega_D}
\end{equation}

We obtain
\begin{equation}
{1 \over \tau}{dE(\tau) \over d\tau} = -{2e \over \tau^2} \sum_{s=1}^{4}\int
 [d{\bf k}]
({\bf k}^2_{\perp} +m^2)
\lambda_{s}\vert f_{{\bf k}s}^{+}\vert ^2 ,
\label{boost_MaxD3}
\end{equation}

We would like to compare the field theory calculation with a semiclassical
transport approach which uses as a source of particle production Schwinger's
production rate for time independent fields.

Assuming boost-invariant initial conditions for $f$, invariance
of the Boltzmann-Vlasov assures that the distribution function
is a function only of the boost invariant variables ($\tau,\eta-y$)
or ($\tau,p_\eta$). The kinetic equation reduces to
\begin{eqnarray}
{\partial f \over \partial
\tau}+eF_{\eta\tau}(\tau){\partial f \over
\partial p_{\eta} }
&=&\pm [1\pm 2f({\bf{p}},\tau)] e\tau \vert E(\tau)\vert \nonumber \\
&& \times \ln {\left [1 \pm \exp{\left (- {\pi
(m^2+{\bf{p}}^2_{\perp}) \over e\vert E(\tau)
 \vert }\right )}\right ]}\delta (p_{\eta}).\nonumber \\
\label{boost_BV}
\end{eqnarray}

Turning now to the Maxwell equation, we have  that
\begin{eqnarray}
-\tau {dE \over d\tau}=j_{\eta}=j^{cond}_{\eta}+j^{pol}_{\eta}\,,
\label{boost_BVmax}
\end{eqnarray}
where $j^{cond}$ is the conduction current and
$j^{pol}_\mu$ is the polarization current due to pair creation$^9$.

\centerline{\epsfysize=6in \epsffile{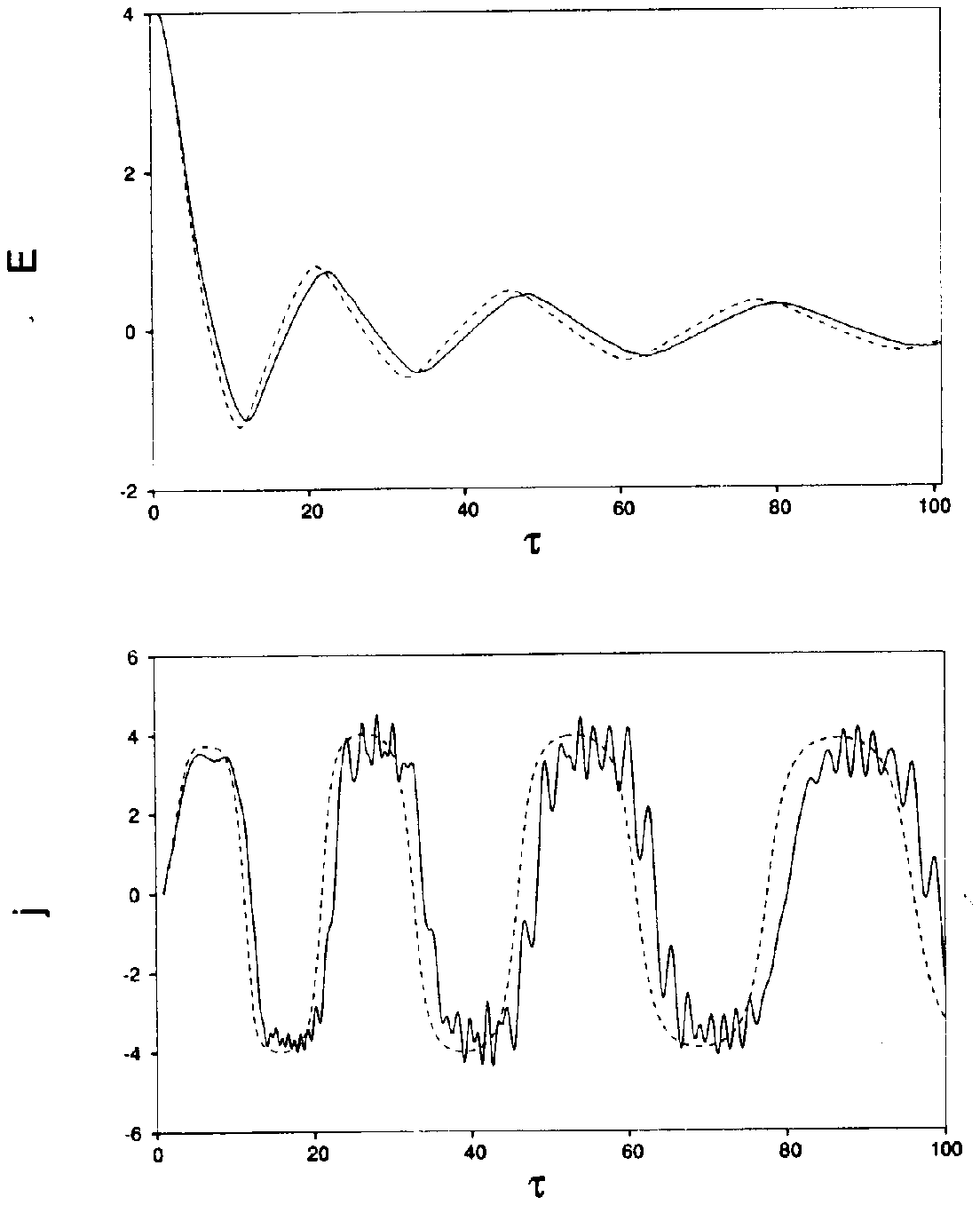}}\vskip4pt\centerline{\bf Fig.
2}
\hangindent\parindent{Proper time evolution of the electric field $E(\tau)$ and
the fermion current $j_{\eta}(\tau)$ for initial conditions$ E(\tau = 1) =
4.0$.
The field theory calculation is compared to the semiclassical transport
approach with a Schwinger source term.}\\ \\

Thus in (1+1) dimensions we have
\begin{eqnarray}
j^{cond}_{\eta}&=&2e\int {dp_{\eta} \over 2\pi \tau p_{\tau}}p_{\eta}
f(p_{\eta},\tau) \nonumber \\
j^{pol}_{\eta}&=&{2 \over F^{\tau\eta}}\int {dp_{\eta} \over
2\pi \tau p_{\tau}} p^{\tau}{D f \over D \tau} \nonumber\\
&=& \pm [1\pm 2f(p_{\eta}=0,\tau)]{m e\tau \over \pi}
{\rm sign}[E(\tau)] \ln {\left [1\pm \exp{\left (-{\pi m^2 \over
 \vert eE(\tau) \vert} \right )}\right ]}.\nonumber \\
\label{boost_jpol}
\end{eqnarray}

In figure two we compare the results of a numerically solving the back reaction
equations in the field theory with the semiclassical transport approach.
We find that for this approximation (no rescattering of quarks), the
semiclassical
method does reasonable well when compared to a  coarse grained
in time and momentum field theory calculation.\\ \\
{\bf DYNAMICAL EVOLUTION OF A NON-EQUILIBRIUM CHIRAL PHASE TRANSITION} \\

Recently there has been speculation that disoriented chiral condensates (DCC's)
can be formed following a heavy ion collision, and these condensates, formed
during a quenched phase transition from the unbroken high temperature phase,
could lead to events with a nonequilibrium distribution of charged to neutral
pions. To see whether these ideas made sense we studied numerically the time
evolution of pions  produced following a heavy ion collision using the linear
sigma model, starting from the unbroken phase.  The quenching (if present) in
this model is due to the expansion of the initial Lorentz contracted energy
density by free expansion into vacuum.
  Starting from an approximate equilibrium configuration at an initial proper
time $\tau$ in the disordered phase we study the transition to the ordered
broken symmetry  phase as the system expands and cools. We determined$^3$ the
proper time evolution of the effective pion mass, the order parameter
$<\sigma>$
 as well as the pion two point correlation function. We studied the phase
space of initial conditions that lead to instabilities (exponentially growing
long wave length modes) which can lead to disoriented chiral condensates.
The model we used to study the chiral phase transition is the
$\sigma$ model in the large $N$ approximation. This model has the
correct chiral properties and gives a reasonable description of low energy
pion dynamics.
The $O(4)$ $\sigma$ model is described by the Lagrangian:
 \bq
L= \{{1\over 2} \partial\Phi \cdot \partial\Phi - {1\over 4}
\lambda (\Phi \cdot \Phi - v^2)^2 + H\sigma\}.
\eq
The mesons  form an $O(4)$ vector
$$\Phi = (\sigma, \pi_i) \hspace{.1in}$$
Introducing the order parameter: $$ \chi = \lambda (\Phi \cdot \Phi-v^2).$$
we have the alternative Lagrangian:
\bq
L_2 = -{ 1 \over 2} \phi_i (\Box + \chi) \phi_i + {\chi^2 \over 4 \lambda} +
{1 \over 2} \chi v^2 + H \sigma
\eq

Perform the Gaussian path integral over the $\Phi$ field. Evaluate
the remaining $\chi$ integral at the stationary phase point .
Legendre transforming:
$$
\Gamma[\Phi,\chi] = \int d^4x[ L_2(\Phi,\chi,H) + { i \over 2}N {\rm tr~ln}
G_0^{-1}]
$$
$$
G_0^{-1}(x,y) = i[\Box + \chi(x)] ~\delta^4(x-y)
$$
$$
[\Box + \chi(x)] \pi_i = 0 ~~~~ [\Box + \chi(x)]\sigma = H
$$

\bq \chi= - \lambda v^2 + \lambda (\sigma^2 + \pi \cdot \pi) + \lambda
N  G_0 (x,x).
\eq
As in the quark-gluon plasma problem we consider the kinematics of an
ultrarelativistic Heavy Ion Collision which possesses longitudinal  Boost
invariance as
the center of mass energy goes to infinity. Energy densities become function of
the
proper time only.
Natural coordinates
are the
proper time $\tau$ and the spatial rapidity
$\eta$ defined as
\bq \tau\equiv(t^2-x^2)^{1/2}, \qquad \eta\equiv{1\over 2}
\log({{t-x}\over{t+x}}).
\eq
We assume that the mean  (expectation) values of the fields     $\Phi$  and
$\chi$ are functions
of $\tau$ only (homogeneity in the constant $\tau$ hypersurface)

\ba
\tau^{-1}\partial_\tau\ \tau\partial_\tau\ \Phi_i(\tau)
 +\ \chi(\tau)\ \Phi_i(\tau) &=&
 H \delta_{i1} \nonumber \\
\chi(\tau) &=&\lambda\bigl(-v^2+\Phi_i^2(\tau)+
N <\phi^2(x,\tau)>\bigr).
\ea
On the otherhand the fluctuation fields (which are quantum operators) are
functions of both $x$ and $t$ and obey the sourceless equation:
\bq
\Bigl(\tau^{-1}\partial_\tau\ \tau\partial_\tau\ - \tau^{-2}\partial^2_\eta
-\partial^2_\perp + \chi(x)\Bigr)
\phi(x,\tau)=0.
\eq

\bq
    G_0 (x,y;\tau) \equiv < \phi(x,\tau) ~\phi(y,\tau)>.
\eq
We expand this field in an orthonormal basis:
\bq
\phi(\eta,x_\perp,\tau)\equiv{1\over{\tau^{1/2}}} \int \ddk\bigl(\exp(ik x)
f_\kk(\tau)\ a_\kk\ + h.c.\bigr)
\eq
where $k x\equiv k_\eta \eta+\vec k_\perp \vec x_\perp$, $\ddk\equiv dk_\eta
d^2k_\perp/
(2\pi)^3$ and the mode functions $f_\kk(\tau)$ evolve according to (a dot here
denotes the derivative with respect to the proper time $\tau$):
\bq
\ddot f_\kk +
\bigl({k_\eta^2\over{\tau^2}}+\vec k_\perp^2 + \chi(\tau) +
{1\over{4\tau^2}}\bigr)
f_\kk=0. \label{finaleqmode}
\eq
\bq
\chi(\tau) = \lambda\Bigl(-v^2+\Phi_i^2(\tau)+{1\over \tau}N \int \ddk
|f_\kk(\tau)|^2\  (1+2\ n_\kk) \Bigr).\label{eq:chi}
\eq

If we assume the initial density matrix is one of local thermal equilibrium
then we have at $\tau=\tau_0$ (the surface
of constant energy density and temperature $T_0$) that:
$$
n_k = {1 \over e^{\beta_0  E^0 _k} -1}
$$
where $ \beta_0 = 1/T_0$ and $E^0_k= \sqrt{k^2+\chi(\tau_0)}$.

In choosing initial conditions we assumed that the initial value of
$\chi$ was determined by the equilibrium gap equation for an initial
temperature
of $ 200 MeV$. The phase transition in this model occurs at a critical
temperature
of $160 MeV$. We chose initial $\sigma$ and $\pi_i$ expectation values
consistent
with the constraint
\bq \vec\pi^2(\tau_0)+\sigma^2(\tau_0)=\sigma^2_T
\eq
where $\sigma_T$ is the
equilibrium value of $\Phi$ at the initial temperature $T$.
which we choose to be a  temperature of $200 MeV$. We
varied the value of the initial proper time derivative of the sigma field
expectation value and found that there is a narrow range of initial values that
lead to the growth of instabilities. Namely
\bq  .25 <  \vert \dot{\sigma} \vert   < 1.3
\eq
 Surprisingly when $ \vert \dot{\sigma} \vert > 1.3 $ instabilities no longer
occur.

Figures 3-4 summarize the results of the numerical simulation for the
evolution of the system (\ref{eq:chi})--(\ref{finaleqmode}).
We display the auxiliary field $\chi$ in units of $fm^{-2}$  , the classical
fields $\Phi$ in units of $fm^{-1}$ and the proper time
in units of $fm^{-1} $   ($ 1fm^{-1} = 197
MeV$).
In Fig.~3 the proper time evolutions of the auxiliary field $\chi$
field is presented for four different initial conditions.

We are interested in knowing how our results differ from the case where
the system evolves in local thermal equilibrium which is described by two
correlation lengths, the inverse of the effective pion mass associated with
$\chi$, and the inverse of the proper time evolving effective temperature
$T(\tau) = T_{0}({\tau_{0}\over \tau})^{\frac{1}{3}}$ discussed earlier.  We
see from Fig. 4 that in the case that $\sigma(1) = \sigma_{T}$, $\pi^{i}(1) =
0$ and $\dot{\sigma}(1) = -1$, where maximum instability exists, complex
structures are formed as contrasted to the local thermal equilibrium evolution.
 The interpolating phase space distribution $n(k_{\eta}, {\bf k}_{\perp},
\tau)$ obtained numerically, clearly exhibits a larger correlation length in
the transverse direction than the equilibrium one and has correlation in
rapidity of the order of 1-2 units of rapidity.  We notice that in both
directions there is structure which does not lend itself to a simple
interpretation.  On the other hand the local thermal equilibrium evolution is
quite regular apart from the normalization of the
distributions that are changing with time due to oscillation in the quantity
$\chi(\tau)$ which is damped to its equilibrium value once the system expands
sufficiently.\\ \\
{\bf Acknowledgments }\\

The work presented here was done in
collaboration with Emil Mottola, Yuval Kluger, Salman Habib, Juan Pablo Paz,
Ben Svetitsky, Judah Eisenberg, Paul Anderson and John Dawson.
This work was supported by the Department of Energy.

\newpage
\centerline{\epsfysize=7.2in \epsffile{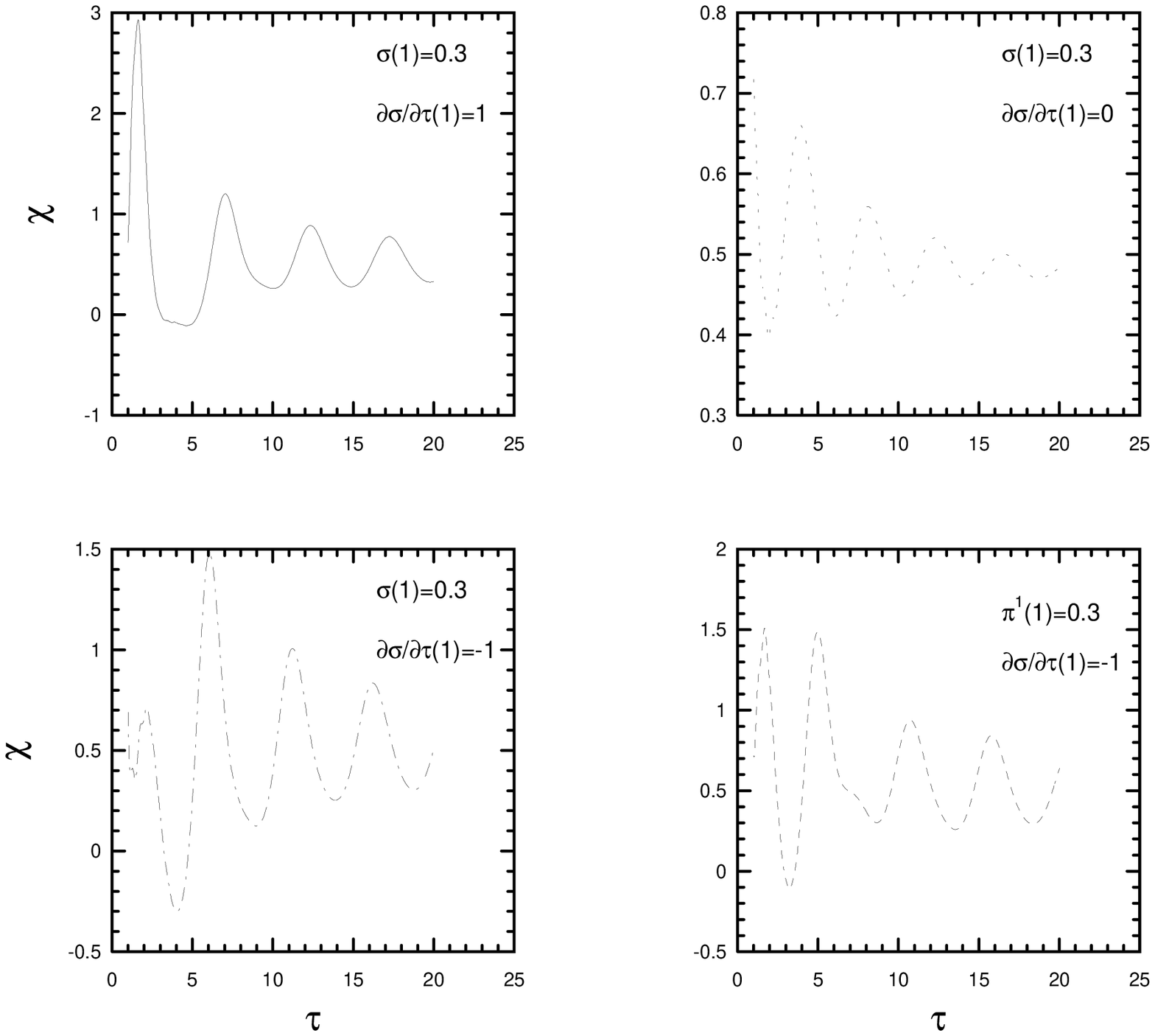}}\vskip4pt \centerline{\bf
Fig. 3}
\hangindent\parindent{Proper time evolution of the $\chi$ field for four
different initial conditions with $f_{\pi} = 92.5 MeV$.}\\ \\
\centerline{\epsfysize=6.7in \epsffile{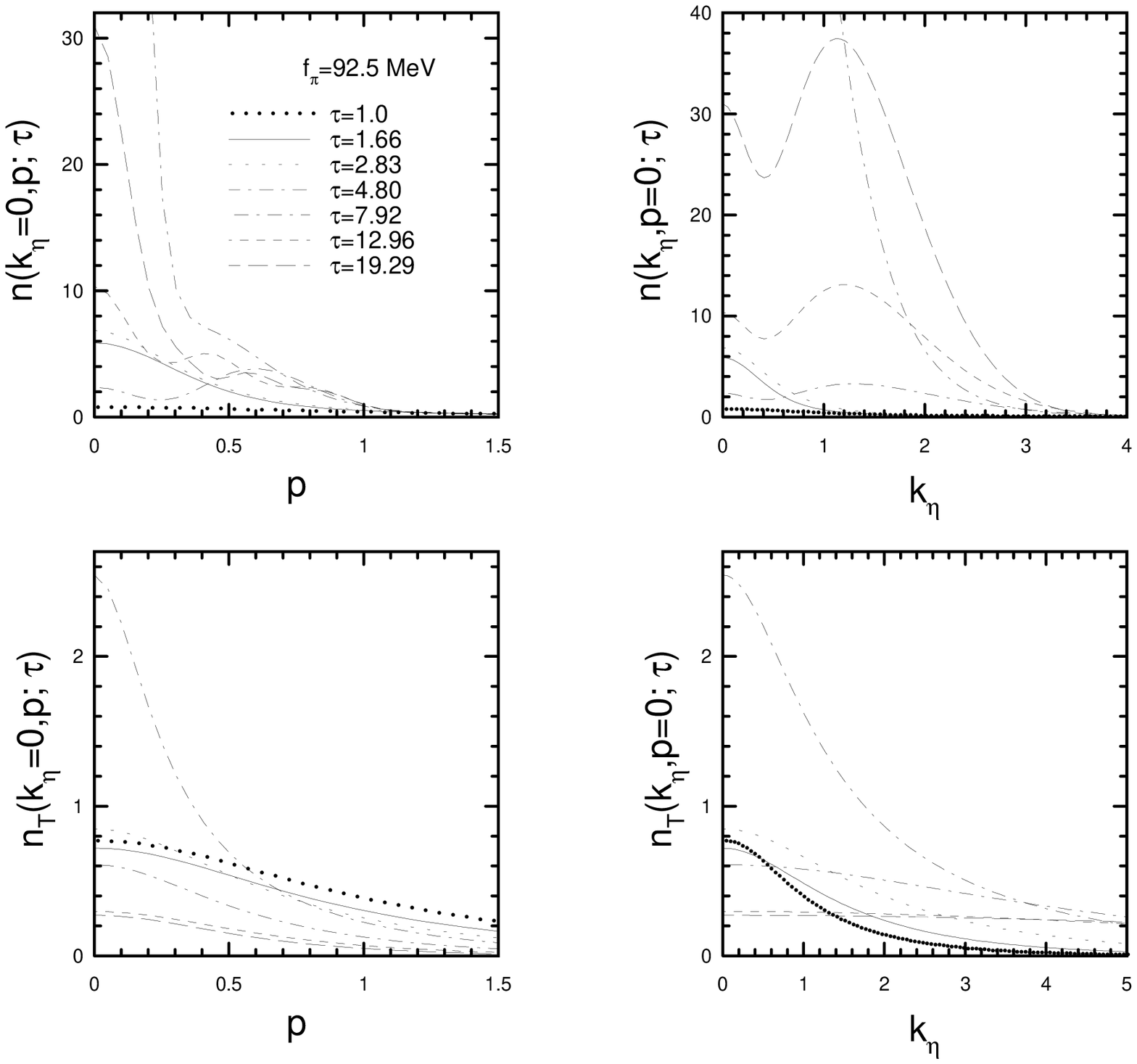}}\vskip4pt\centerline{\bf Fig.
4}
\hangindent\parindent{Slices of $k_{\eta} = 0$ and $p \equiv \vert{\bf
k}_{\perp}\vert = 0$ of the proper time evolution of the interpolating phase
space particle number density $n(k_{\eta} , {\bf k}_{\perp} , \tau)$ for
$\sigma(1) = \sigma_{T}$, $\pi^{i}(1) = 0$ and $\dot{\sigma} (1) = -1$ compared
with the corresponding local thermal equilibrium densities $n_{T}(k_{\eta} ,
{\bf k}_{\perp} , \tau)$.} \\ \\
\end{document}